\documentclass[12pt]{article}       
\usepackage{amsmath}
\usepackage{graphicx}
\usepackage{authblk}
    
\graphicspath{ {./media/} }
\usepackage{subcaption}
\usepackage{url}
\usepackage{rotating}   
\usepackage{xspace} 
\usepackage{hyperref}
\usepackage{array}

\newcommand{\altsens}{BRS-T360\xspace} 
\newcommand{\Ry}{$\theta$ }
\newcommand{\htwo}{$H_{2}$ }
\newcommand{\hinf}{$H_{\infty}$ }

\newcolumntype{L}[1]{>{\raggedright\arraybackslash}m{#1}}

\DeclareMathOperator*{\argmin}{arg\,min}

\title{Multi-scale optimal control for Einstein Telescope active seismic isolation}

\author[1,2]{Pooya Saffarieh}
\author[1,2,3]{Nathan A. Holland}
\author[1,2]{Michele Valentini}
\author[1,2]{Jesse van Dongen}
\author[1,2,4]{Alexandra Mitchell} 
\author[1]{Sander Sijtsma} 
\author[1,2]{Armin Numic}
\author[5]{Wouter Hakvoort}
\author[1,2]{Conor Mow-Lowry}
\affil[1]{Dutch National Institute for Subatomic Physics, Nikhef, Amsterdam, The Netherlands}
\affil[2]{Vrije Universiteit Amsterdam, Amsterdam, The Netherlands}
\affil[3]{LIGO Laboratory, California Institute of Technology, Pasadena, USA}
\affil[4]{Stanford University, Stanford, USA}
\affil[5]{University of Twente, Enschede, The Netherlands}

\begin{document}     

\maketitle


\begin{abstract} 


We present a multi-scale optimal control framework for active seismic isolation in the Einstein Telescope, a third-generation gravitational-wave observatory. Our approach jointly optimizes feedback and blending filters in a cross-coupled opto-mechanical system using a unified cost function based on the “acausal optimum,” which quantifies sensor signal-to-noise ratios across frequencies. This method enables efficient re-optimization under varying sensor configurations and environmental conditions. We apply the framework to two candidate sensing systems using their modeled sensitivity: OmniSens—a six-degree-of-freedom inertial isolation system—and BRS-T360, which combines Beam Rotation Sensor (BRS) as an inertial tilt sensor with T360 as a horizontal seismometer. We demonstrate superior low-frequency isolation with OmniSens, reducing platform motion by up to two orders of magnitude near the microseism. The framework allows for ready optimization and projection of sensor noise to metrics relevant to the performance of the instrument, aiding the design of the Einstein Telescope.

\end{abstract}

\section{Introduction}

The optimal design of control filters is essential within the complex system of a gravitational wave (GW) detector, which require extreme precision to detect weak astrophysical signals. These detectors comprise multiple hierarchical subsystems—such as seismic isolation, suspension, and cavity control—each with its own set of interacting components spanning various mechanical degrees of freedom (DoF). Within this complex architecture, control filters are essential for shaping system dynamics in a way that minimizes the influence of disturbances on the detector’s most sensitive measurement channels.

Current GW detectors are limited by low-frequency noise dominated by a combination of seismic noise and control noise, below 3Hz \cite{collaborationAdvancedLIGO2015, acerneseAdvancedVirgoSecondgeneration2014,capoteAdvancedLIGODetector2025,acerneseVirgoDetectorCharacterization2023}. The cross couplings---especially tilt-to-translation coupling \cite{matichardReviewTiltFreeLowNoise2015}---play a major role in reducing the sensitivity at low frequency.

Achieving improved low-frequency sensitivity is essential for observing a broader range of astrophysical events over longer timescales, including mergers involving stellar-mass and intermediate-mass black holes. These systems are key to advancing our understanding of fundamental physics and validating diverse theoretical models \cite{maggioreScienceCaseEinstein2020}. The Einstein Telescope (ET) is a proposed third-generation gravitational wave observatory \cite{punturoEinsteinTelescopeThirdgeneration2010,hildSensitivityStudiesThirdgeneration2011}. It is designed to surpass the capabilities of current second-generation detectors-particularly in the low-frequency regime, where seismic and control noise are dominant limiting factors \cite{collaborationAdvancedLIGO2015, acerneseAdvancedVirgoSecondgeneration2014, capoteAdvancedLIGODetector2025, acerneseVirgoDetectorCharacterization2023}.
Addressing this low-frequency noise barrier requires significant upgrades, including enhancements to active isolation systems aimed at suppressing optical component motion at low frequencies \cite{mow-lowry6DInterferometricInertial2019}. However, designing and modelling a high-performance active isolator is a time-intensive task, especially when accounting for realistic environmental inputs and critical tilt-to-translation coupling effects \cite{matichardReviewTiltFreeLowNoise2015}. Control filter design is inherently dependent on the dynamics of sensors, disturbances, and plant. Any modification to these elements necessitates updating control filters to maintain optimal feedback performance and consequently updating the model of the noise limited sensitivity of the detector.

Optimal control has previously been investigated in the design of blending filters for GW seismic isolation platforms \cite{Thomas2019}. Building on this, an improved weighting function and a generalized plant formulation were later introduced for KAGRA’s active isolation system, incorporating both seismic and sensor noise dynamics \cite{tsangOptimizingActiveSeismic2025}. However, in that work, the blending filter is treated independently from the feedback control loop in which it is implemented. Since the blending filter is nested within the feedback loop, this separation implies that the blending filter may be optimal in isolation, while the overall control loop is not globally optimal. In addition, due to the optimization algorithm used, the complexity of the optimal controller produced in these works is in the same order as the plant, making it infeasible to scale to larger plants with cross-couplings and noise dynamics included. A multi-objective optimization approach for the ET suspension system has recently been investigated to identify optimal combinations of actuation filters for the final stage of the suspension, while also quantifying the contribution of Digital to Analog Converter (DAC) noise to the main interferometer axis \cite{Sander2025}. This study employs a non-smooth gradient descent algorithm and adopts a more flexible framework that simultaneously considers multiple design criteria, including robustness. It advances optimal control techniques within the context of gravitational wave detectors. Given that control loops often interact, it is crucial to design filters for one loop while accounting for the impact of cross-couplings on the performance of others.

In this paper, we formulate the control of an active isolation platform as a multi-scale optimal-control problem that allows rapid testing and re-optimization for differing input motions and sensor configurations considering cross-couplings and nested loops. The nested loop structure naturally introduces multiple scales, and by optimizing all levels simultaneously, the problem becomes inherently multi-scale. At the core of the proposed method is the so-called "acausal optimum", denoted by $\xi$, which is formed by taking the minimum across all sources of noise. Its inverse acts as a normalization factor for the platform motion, converting the closed-loop residual motion into a Signal-to-Noise Ratio (SNR). This normalization factor then enforces the optimizer to use sensors in regions with high SNR and penalizes the loop gain where SNR is low. We construct optimized noise curves for two configurations of sensors: OmniSens \cite{mow-lowry6DInterferometricInertial2019}, a novel 6D inertial isolation system; and a combination of Nanometrics T360 seismometers \cite{T360}, and Beam Rotation Sensor (BRS) \cite{venkateswaraHighprecisionMechanicalAbsoluterotation2014} upgraded with interferometric readout \cite{mitchellIntegrationHighperformanceCompact2024} here called \altsens. This not only sets the stage for further noise propagation and suspension design studies for ET but also provides a novel approach for tackling multi-scale optimization problems abundant within the complex systems of a gravitational wave observatory.

The structure of the paper is as follows: Section 2 introduces two active isolation configurations that serve as models for applying the proposed multi-scale optimization algorithm. In Section 3, we detail the derivation of the transfer functions necessary for problem formulation, enabling the application of optimization to the specified configurations. This section also analyzes the role of the weighting function and how the interplay between the blending filter and controller introduces distinct scales within the loop architecture. Section 4 presents the optimization results in terms of noise budgets. Additional technical details, including the derivation of the equations of motion and the optimized control filters, are provided in the appendices.

\section{Active isolation configurations}

\begin{figure}
  \centering
  \begin{subfigure}{.47\textwidth}
    \centering
    \includegraphics[width=1\linewidth]{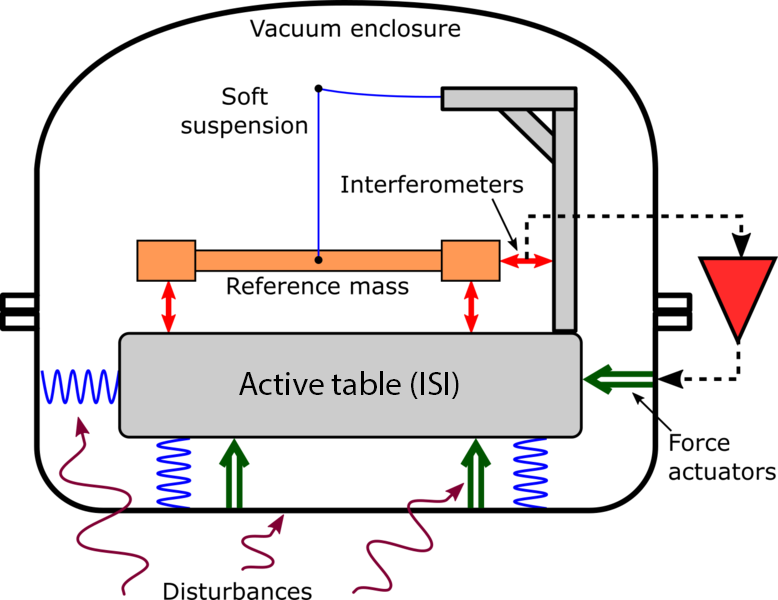}
    \caption{ OmniSens}
    \label{fig:omnisens-diagram}
  \end{subfigure}%
  \hspace{0.05\textwidth} 
  \begin{subfigure}{.47\textwidth}
    \centering
    \includegraphics[width=1\linewidth]{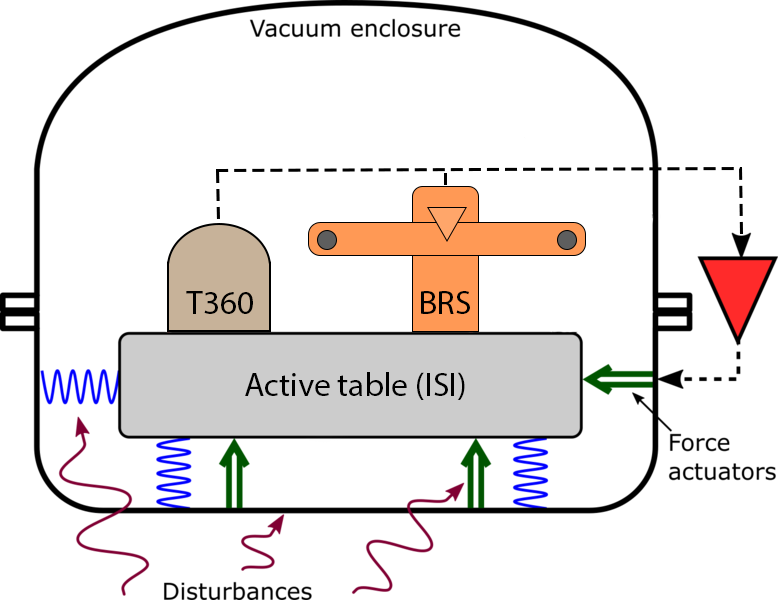}
    \caption{ \altsens}
    \label{fig:altsens-diagram}
  \end{subfigure}
  \caption{ Active platform with two inertial sensor configurations. The left figure shows the OmniSens configuration using a single reference mass for rotation and translation. The right figure shows the \altsens configuration using separate rotation and translation sensors. Both configurations use interferometric (HoQI) sensors to track the relative displacement of the platform.}
  \label{fig:active-platform-configs}
  \end{figure}

The active platform comprises two primary components: the Internal Seismic Isolation (ISI) table and inertial sensors. The ISI is equipped with six actuators and six displacement sensors, enabling sensing and actuation across all rigid degrees of freedom. The ISI model used in this work is obtained from \cite{LIGO_G070156} with modifications to stiffness and elasticity parameters as detailed in Appendix B.

Since for any inertial sensor measuring horizontal motion tilt couples to translation at low frequency \cite{matichardReviewTiltFreeLowNoise2015} we need to have both rotation and translation sensing. 

The combination of these sensors and the active table are parts of a feedback loop which suppresses the ground disturbances on the platform. You can see a schematic of these configurations in figure \ref{fig:active-platform-configs}.

OmniSens consists of a softly suspended reference mass from a silica fibre. This combines with a 6 DoF interferometric sensor reading of this test mass. Since the system is decoupled and symmetric, we can decompose the dynamic on two orthogonal 2-dimensional harmonic oscillators and analyze it. The 2 DoF model captures almost all the complexity of the 6 DoF platform. For the reference mass, a derivation of the equation of motion and thermal noise is presented in Appendix A. 

In this paper, we modelled three sources of noise: thermal noise, Homodyne Quadrature Interferometer (HoQi) displacement sensor noise, and ground noise. For HoQi noise we fit a transfer function to empirical measurements obtained from \cite{mitchellIntegrationHighperformanceCompact2024}. The ground noise is obtained by using the GIGS sensor, at The Laboratori Nazionali del Gran Sasso, Italy (LNGS), between April 2022 and December 2023---to capture a high-noise scenario the 90th percentile from December 2023 is used. Seismic data from LNGS is freely accessible through the FDSN service of the Italian national seismic network \cite{INGV_RSN_2005}.

In \altsens configuration, the reference mass is replaced with an upgraded BRS sensor for rotational sensing and a T360 seismometer for horizontal motion sensing. These sensors were chosen for their established reliability and widespread use in the gravitational wave community, allowing for a meaningful comparison of OmniSen's performance against other alternatives.

The acceleration noise performance of the T360 seismometer is directly available from the manufacturer \cite{T360}---we convert it into an amplitude spectral density with units of nm / s$^{2}$ $\sqrt{\text{Hz}}$. 

The tilt noise performance for the upgraded BRS is constructed from the readout noise and thermal noise of the flexure which gives its proof mass low-frequency tilt compliance. Only these noise sources are considered to provide a fair comparison with the OminSens configuration, despite other limitations to its performance existing in realised implementations to date \cite{venkateswaraHighprecisionMechanicalAbsoluterotation2014,LIGO_BRS_TiltSubtraction_2017,Windproofing_LIGO_2020,RossMP_PhDthesis_2020}. Upgrading the readout of the BRS to interferometric sensors \cite{mitchellIntegrationHighperformanceCompact2024} is motivated by the successful inclusion of the same in its more compact sibling sensor the Cylindrical Rotation Sensor \cite{rossVacuumcompatibleCylindricalInertial2023}.

\begin{table}[t]
\centering
\setlength{\tabcolsep}{5pt} 
\renewcommand{\arraystretch}{1.2} 
\begin{tabular}{|L{0.15\textwidth} | L{0.3\textwidth} |L{0.15\textwidth} | L{0.3\textwidth}|}
\hline
\textbf{Symbol} & \textbf{Description} & \textbf{Symbol} & \textbf{Description} \\
\hline \hline
$G(s)$ & Plant transfer function & $\xi(f)$ & Acausal optimum \\ \hline
$K(s)$ & Control filter & $S(s)$ & Sensitivity function $S(s) = \frac{1}{1 + K(s) G(s)}$ \\ \hline
$H(s)$ & Blending filter (high-pass) & $P(s)$ & Generalized plant \\ \hline
$\bar{H}(s)$ & Complementary blending filter (low-pass) & $W_{p}(s)$ & Weighting function \\ \hline
$\eta_{\rm dist}(s)$ & Disturbance noise coloring filter & $w(s)$ & White noise input \\ \hline
$\eta_{\rm disp}(s)$ & Displacement sensor noise coloring filter & $\eta_{\rm sens}(s)$ & Inertial sensor noise coloring filter \\ \hline
\end{tabular}
\caption{Overview of the commonly used symbols in the paper. Subscripts indicate the degree of freedom (DoF): $\theta$ for rotational loop and $x$ for horizontal loop. For example, $K_{\theta}$ and $\eta_{\theta,\mathrm{dist}}$ correspond to the \Ry loop, while $K_{x}$ and $\eta_{x,\mathrm{dist}}$ correspond to the x loop.}
\label{tab:variables_definition}
\end{table}

\section{Optimal control} 

\begin{figure}[t]
\centering
\includegraphics[width=1\textwidth]{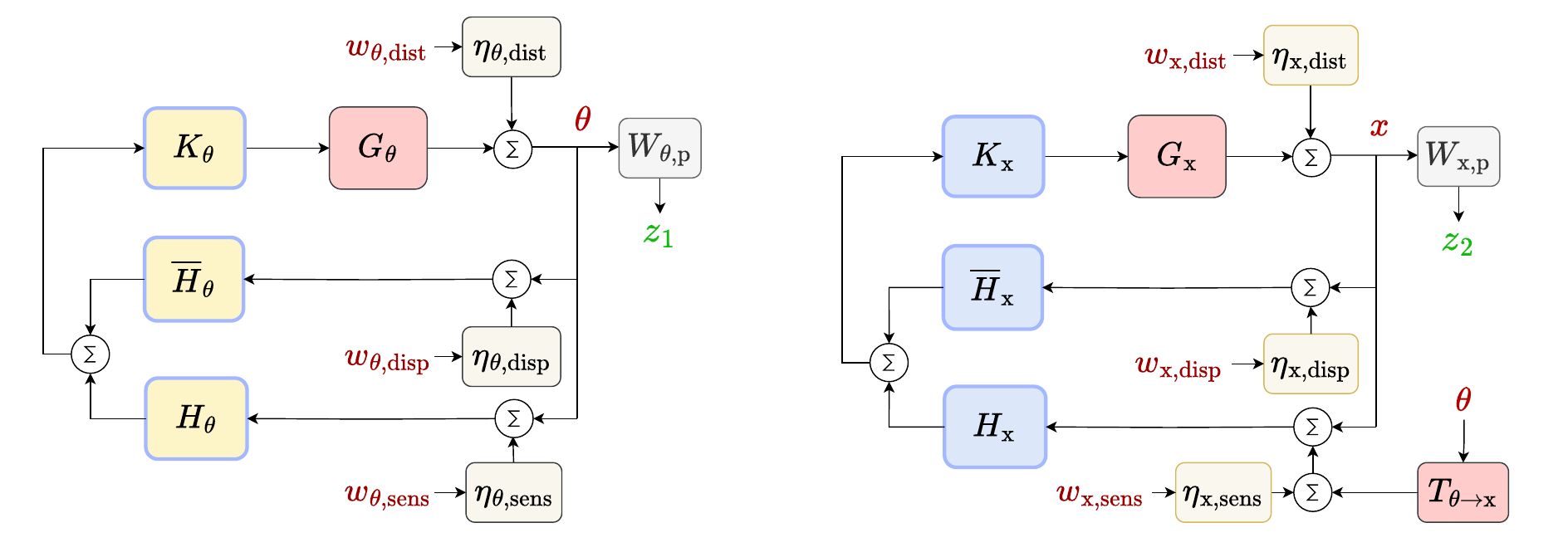}
\caption{Control schemes for rotation (left, $\theta$) and horizontal translation (right, $x$). White noise inputs $w$ are shaped by noise coloring filters $\eta$ to model real noise sources. The transfer function $T_{\theta \to x}$ captures coupling from rotation to translation. Weighting functions $W$ normalize the closed-loop residual motions $\theta$ and $x$ in order to produce performance outputs $z_1$ and $z_2$. $G_{\theta}$ and $G_x$ are the plant transfer functions, while $K_{\theta}$ and $K_x$ are the feedback controllers for rotation and translation, respectively. $H_{\theta}$ and $H_x$ are the blending filters, with complementary filters $\bar{H}_{\theta}$ and $\bar{H}_x$. $\eta_{\mathrm{dist}}$, $\eta_{\mathrm{disp}}$, and $\eta_{\mathrm{sens}}$ represent the disturbance, displacement-sensor, and inertial-sensor noise coloring filters, respectively; the subscript $\theta$ or $x$ indicates the rotational or horizontal loop.}
\label{fig:mimo-scheme}
\end{figure}

\begin{figure}[ht]
  \centering
  \includegraphics[width=1\textwidth]{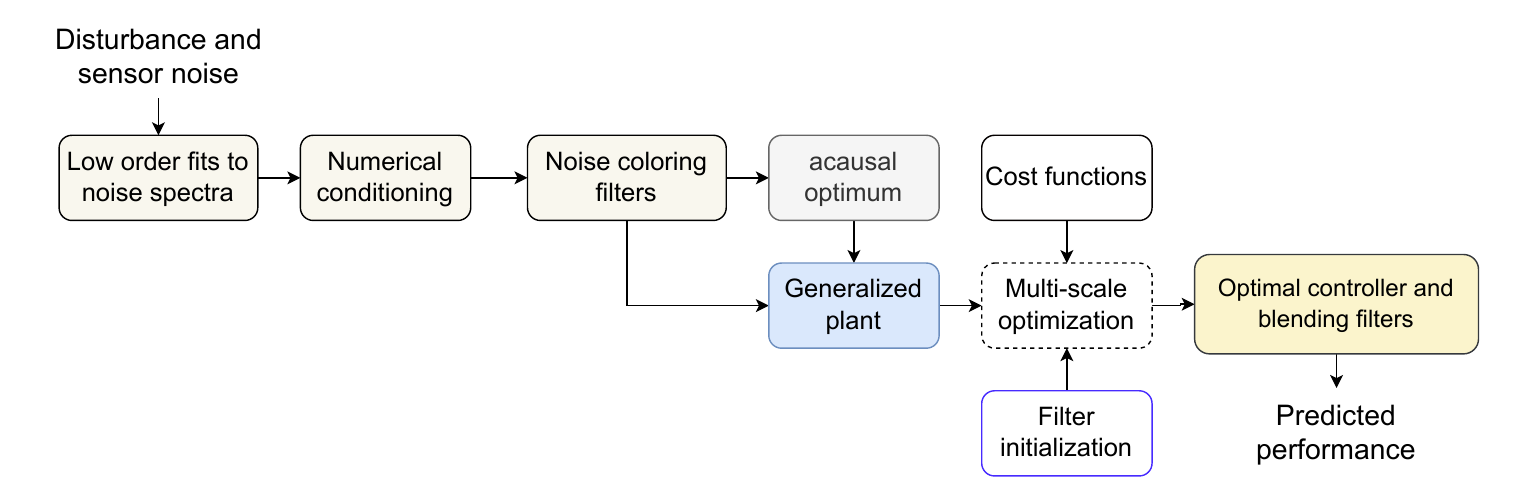}
  \caption{ Procedure for Setting Up the Optimization in the Given Control Problem. The process begins by modeling the noise and transforming it into a format suitable for numerical computation. Using these noise models, the acausal optimum is constructed and integrated into the generalized plant, which represents the multi-scale cross-coupled system. This procedure consistently generates performance predictions based on the input noise spectra.}
  \label{fig:procedure}
  \end{figure}

\begin{figure}[t]
\centering
\begin{subfigure}{.49\textwidth}
  \centering
  \includegraphics[width=1\linewidth]{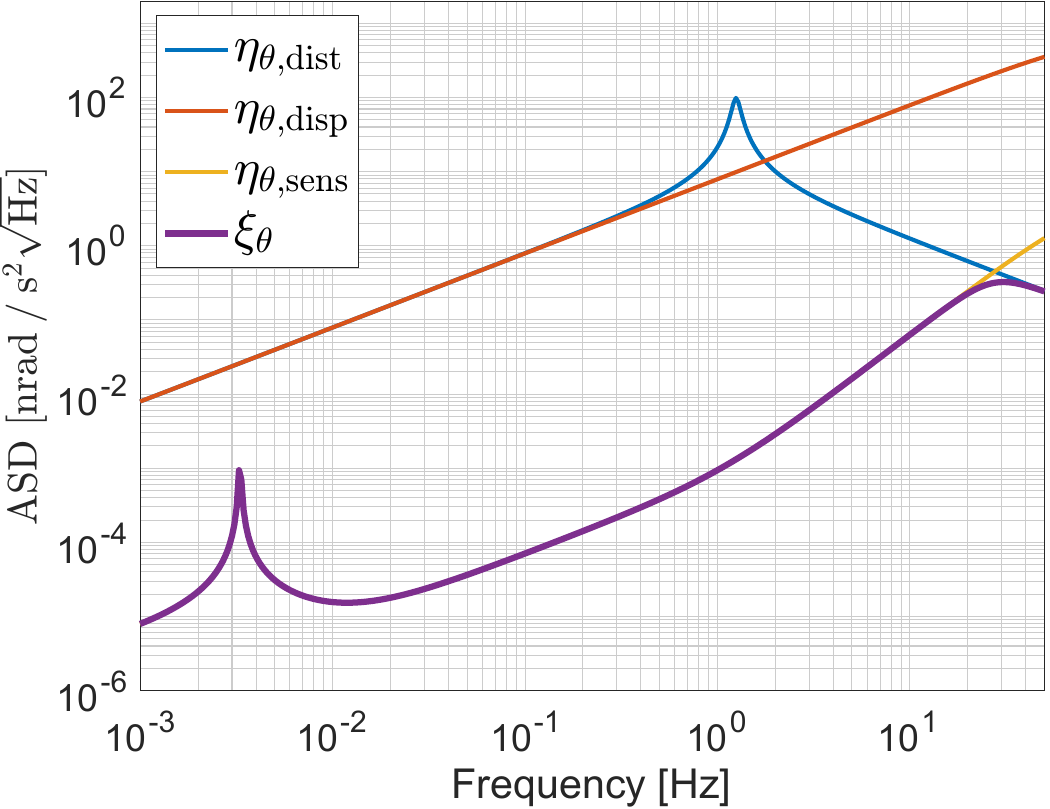}
  \caption{\Ry loop.}
  \label{fig:omnisens-ry-compare}
\end{subfigure}
\begin{subfigure}{.49\textwidth}
  \centering
  \includegraphics[width=1\linewidth]{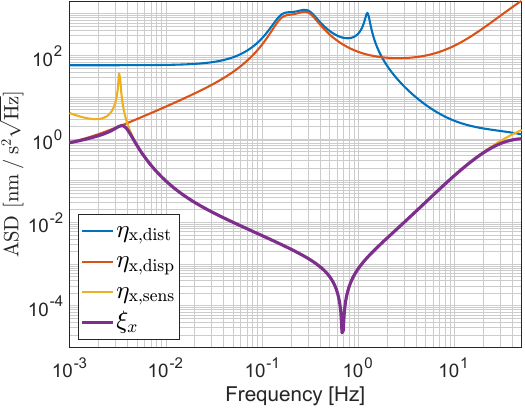}
  \caption{x loop.}
  \label{fig:omnisens-x-compare}
\end{subfigure}
\caption{OmniSens acausal optimum $\xi$ decomposition for both \Ry and x loop. The acausal optimum represents the global minimum across all noise curves, as defined in Eq.~\ref{eq:acausal-optimum}. The curves $\eta_{dist}$, $\eta_{disp}$, and $\eta_{sens}$ captures ground disturbance, displacement sensor noise, and inertial sensing noise, respectively.}
\label{fig:omnisens-idealcurves}
\end{figure}

\begin{figure}[t]
\centering
\begin{subfigure}{.49\textwidth}
  \centering
  \includegraphics[width=1\linewidth]{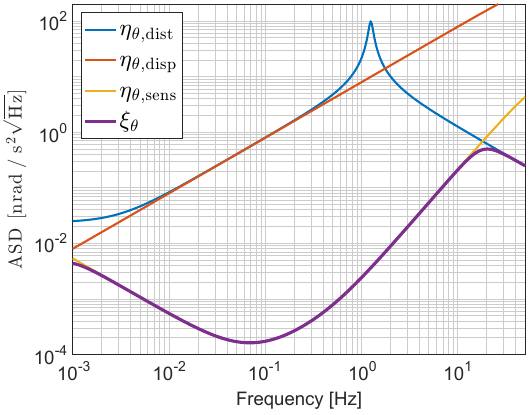}
  \caption{\Ry loop.}
  \label{fig:altsens-ry-compare}
\end{subfigure}
\begin{subfigure}{.49\textwidth}
  \centering
  \includegraphics[width=1\linewidth]{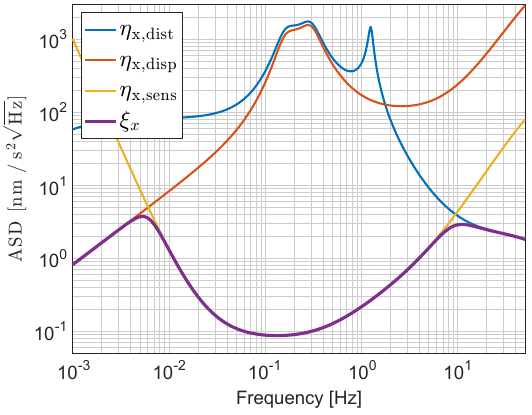}
  \caption{x loop.}
  \label{fig:altesens-x-compare}
\end{subfigure}
\caption{\altsens acausal optimum $\xi$ decomposition for both \Ry and x loop. The curves $\eta_{dist}$, $\eta_{disp}$, and $\eta_{sens}$ captures ground disturbance, displacement sensor noise, and inertial sensing noise, respectively.}
\label{fig:altsens-idealcurves}
\end{figure}

This section presents a multi-scale control design that combines both $H_{\infty}$ (angular) and $H_2$ (translation) norms within a generalized plant framework. Weighting functions derived from the signal-to-noise ratio (SNR) quantify deviation from the acausal optimum---representing the theoretical performance limit---so that optimization targets realistic, sensor-limited performance rather than uniform RMS minimization. In the $H_2$ norm, an additional logarithmic weighting proportional to $f^{-\alpha}$ ($\alpha = 0.5$) ensures equal contribution per frequency decade. Optimization is performed using a subgradient descent algorithm rather than Riccati-based synthesis to enable reduced-order, numerically robust controllers. Although subgradient descent sacrifices strict optimality guarantees, it circumvents the numerical conditioning problems that prevented Riccati methods from converging to true optima in practical, multi-loop implementations.

The core concept of the control system is to implement a feedback loop that blends inputs from various sensors, each offering superior SNR in specific frequency ranges, and feeds this back to the actuators. In frequency bands where the SNR drops below 1, the controller is attenuated and platform motion asymptotes towards its free-running performance. This approach results in platform motion that is quieter than the ground motion. Figure \ref{fig:mimo-scheme} shows the control block diagram for these simulations.

Optimal control assumes all noise inputs are white. Noise coloring filters shape these inputs to reflect realistic disturbances and sensor noise, while weighting functions emphasize specific frequency ranges in the plant output to guide the optimization toward desired performance. The combination of plant, noise filters, and weighting functions forms the generalized plant, which provides the basis for controller optimization. For a detailed introduction to optimal control, see \cite{zhouEssentialsRobustControl1998}.

To formulate the control design as an optimization problem, we follow the procedure shown in Figure \ref{fig:procedure}. Noise inputs are modeled using noise-coloring transfer functions, which define the acausal optimum $\xi$, formally introduced in Equation \ref{eq:acausal-optimum}. These optima are illustrated for both configurations in Figures \ref{fig:omnisens-idealcurves} and \ref{fig:altsens-idealcurves}. Overall, OmniSens demonstrates superior performance owing to its lower thermal noise and increased sensitivity to rotation at low frequencies. In addition, the interferometric readout of the OmniSens reference mass for translation provides improved sensitivity compared to that of the T360.

The weighting functions are constructed as the inverse of these acausal optimums. Together with the plant transfer functions, these form the generalized plant used in the optimization algorithm. Each step is detailed in the following sections.
 
The mechanics are designed to minimize cross-couplings, simplifying and clarifying the control design. We assume the coupling from horizontal translation to rotation is negligible. Noise inputs are modelled as white noise multiplied by noise coloring filters $\eta$. These filters are obtained by fitting transfer functions to each noise source's spectra. The Laplace variable, s, is omitted for notational simplicity. For the tilt disturbance, the noise coloring filter is:

\begin{equation}
  \eta_{\rm \theta, dist} = \tilde{G}_\theta  g_\theta,
\end{equation}

where here $g_{\rm \theta}$ is the transfer function fit to ground motion spectra and $\tilde{G}_{\rm \theta}$ is the transfer function from ground to the platform. Similarly, for the $x$ direction:
\begin{equation}
  \eta_{x,{\rm  dist}} = \tilde{G}_x  g_x \lambda_x,
\end{equation}
where $\lambda_x$ is a low-pass filter with a pole set below table resonance which penalizes the open loop motion, encouraging engagement of feedback at low frequencies. This ensures that the controller maintains high loop gain at low frequencies and provides a summing junction for external inputs, such as observatory drift control signals.

The ISI displacement sensor $d_{\theta}(t)$ observes the ground-platform differential motion:

\begin{equation}
d_{\theta}(t) = \theta(t)l - g_{\theta}(t)l + n(t).
\end{equation}

Where $l$ is the lever arm length from center of rotation and $\theta(t)$ is table tilt. Since the sensor self noise $n(t)$ is uncorrelated from ground rotation, $g_{\theta}(t)$, i.e., $\langle n(t), g_{\theta}(t) \rangle = 0$, the equivalent noise is the quadrature sum of ground and sensor noise which is used in the noise coloring filter, $\eta_{\rm \theta, disp}$. The same procedure applies in the x direction, yielding $\eta_{x, {\rm disp}}$. 

For $\eta_{x,\mathrm{sens}}$ and $\eta_{\theta,\mathrm{sens}}$, which represent the noise coloring filters for the horizontal and tilt inertial sensors respectively, we use transfer function fits, as with other noise coloring filters. 

For implementation, we take the control blocks shown in \ref{fig:mimo-scheme} and rearrange the diagram such that all the white-noise inputs $w$ appear on the left and all performance outputs $z$ on the right. 
We replace complementary filters in the scheme by applying $\bar{H} = 1 - H$. This allows us to only optimize for the $H$ and derive the other by using the complementarity property \cite{tsangOptimizingActiveSeismic2025}. We isolate four filters to be optimized ($K_{x}$, $H_{x}$, $K_{\theta}$, and $H_{\theta}$) and everything which is left is part of the generalized plant shown in \ref{fig:generalized-plant}. We can express the obtained input-output system using a matrix. This matrix can be used in Matlab to define the generalized plant. For the matrix formulation of the generalized plant, see Appendix~C.

\subsection{Loop analysis}

We use a sensor blending for both DoF with the following constraint:
\begin{equation}
  H + \bar{H} = 1,
  \label{eq:blending_constraint}
\end{equation}

where $H$ and $\bar{H}$ are high-frequency and low-frequency filters, respectively.  To understand the effect of weighting and the structure of the loop we can write performance output $z_1$ as a function of all the inputs:

\begin{equation}
        z_1 = W_{\rm \theta,p} ( S_{\theta} w_{\rm \theta, dist} \eta_{\rm \theta, dist}  + S_{\theta} K_{\theta} G_{\theta} w_{\rm \theta, disp} \eta_{\rm \theta ,disp} \bar{H}_{\rm \theta} + S_{\theta} K_{\theta} G_{\theta} w_{\rm \theta disp} \eta_{\theta,sens} H_{\rm \theta} ).  \\
\end{equation}

Here $S_{\theta} = 1 / (1+K_{\theta} G_{\theta})$ is the sensitivity function for \Ry loop.  Using the constraint in equation \ref{eq:blending_constraint}, the fact that all inputs are white noise with unit intensity, and $S_{\theta}+K_{\theta} G_{\theta} S_{\theta}=1$ we can simplify this further to:

\begin{equation}
    z_1 = W_{\theta,p} (  \eta_{\rm \theta, dist} S_{\theta}  + ( \eta_{\rm \theta,disp} (1-H_{\rm \theta}) +  \eta_{\theta, \rm sens} H_{\rm \theta} ) (1-S_{\theta}) ).
\end{equation}

Note the nested structure of the expression above, which features two blending filters. One of these filters is embedded within the primary feedback loop, effectively blending input noise sources with the open-loop response. This layered architecture is why we refer to it as multi-scale optimization, as the inner and outer loops interact in a simultaneous way. The same analysis applies to the x loop as well.

\subsection{Weighting function}

\begin{figure}[ht]
\centering
\includegraphics[width=0.6\textwidth]{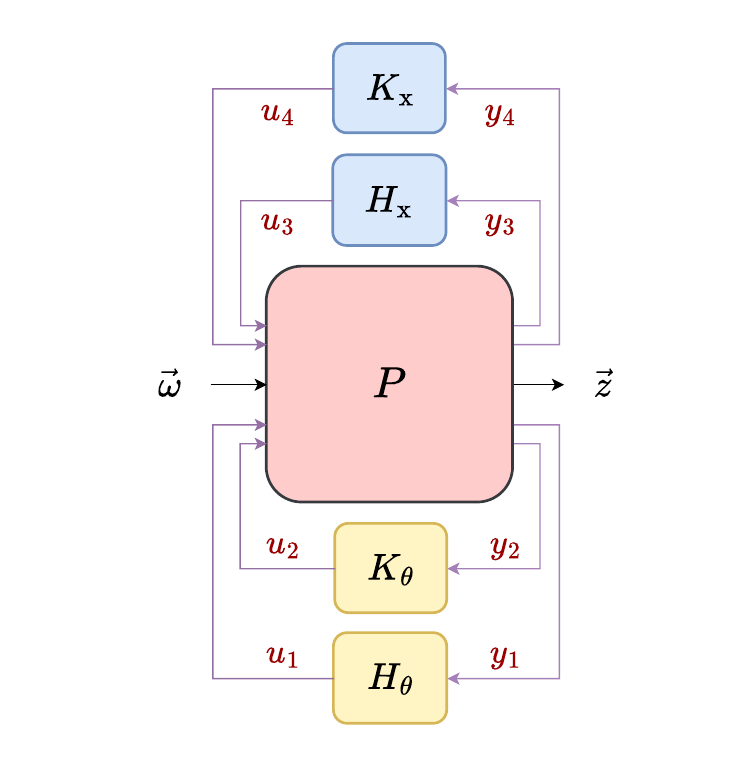}
\caption{ Generalized plant scheme. The generalized plant $P$ includes all transfer functions for noise coloring, weighting filters, and the physical plant. $K_x$ and $H_x$ (control and blending filters for the $x$ loop), and $K_{\theta}$ and $H_{\theta}$ (for the \Ry\ loop) are being optimized. Closed-loop transfer functions from white noise input $\vec{\omega} = [w_{\theta, \rm dist} \ w_{\theta, \rm disp} \ w_{\theta, \rm sens} \ w_{x, \rm dist} \ w_{x, \rm disp} \ w_{x, \rm sens} ]^T$ to output $\vec{z}=[z_1 \  z_2]^T$ are evaluated using various norms to assess optimality.}
\label{fig:generalized-plant}
\end{figure}

In an ideal non-causal case, we can reduce the noise down to the minimum of all sources of noise, namely the acausal optimum. Deviations from this non-causal limit measure the achieved performance with our Linear Time Invariant (LTI) filtering scheme. The acausal optimum for both configurations is shown in Figure \ref{fig:omnisens-idealcurves} and \ref{fig:altsens-idealcurves}. We use this acausal optimum with either $H_2$ or $H_{\infty}$ norm to form the weighting function \cite{zhouEssentialsRobustControl1998}. No weighting results in the lowest RMS platform motion over the full frequency band. However, this makes the controller optimization focus on frequencies with high noise and not exploit the frequency bands where the sensors have their best performance. Thereby, a weighting function is introduced to make the optimization aim at the best performance that is feasible with the used sensors, i.e., improve the relative deviation from the theoretically best performance with the given noise sources. This weighting results in substantially lower closed-loop motion at frequencies with high sensing performance at a negligible RMS penalty.

The acausal optimum, for any of the control blocks depicted in figure \ref{fig:mimo-scheme}, is defined as:

\begin{equation}
    \rm \xi (f) = \left( \left( \frac{1}{\eta_{\rm disp(f)}} \right)^2 + \left( \frac{1}{\eta_{\rm dist(f)}} \right)^2 + \left( \frac{1}{\eta_{\rm sens(f)}} \right)^2 \right)^{-\frac{1}{2}}
\label{eq:acausal-optimum}
\end{equation}

And then the weighting function is defined as:

\begin{equation}
    W_p(f) =  \xi(f)^{-1}
\end{equation}

The weighting function in the Laplace domain, $W_p(s)$, is obtained by fitting a transfer function to the frequency-domain representation, $W_p(f)$. Additionally, to ensure uniform emphasis across all frequency decades in the \htwo norm (on a log-log plot), we apply an extra frequency-dependent weight of the form $f^{-\alpha}$. Setting $\alpha = 0.5$ (the value used in this study) puts the same emphasis across different frequencies. Choosing $1 >\alpha > 0.5$ emphasizes low frequencies, while $0< \alpha < 0.5$ emphasizes high frequencies. To implement the resulting fractional-order transfer function, we use the Oustaloup recursive filter method \cite{oustaloupFrequencybandComplexNoninteger2000}. The complete generalized plant—incorporating all transfer functions, noise coloring filters, and weighting functions—is shown in Figure \ref{fig:generalized-plant}.

\subsection{Cost function}

\begin{figure}[t]
  \centering
  \includegraphics[width=0.5\textwidth]{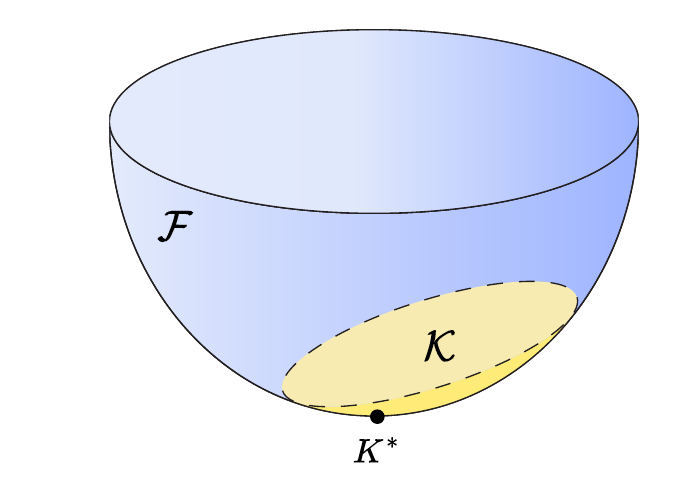}
  \caption{ A graphical representation of optimization. $\mathcal{F}$ is the set of all controllers which make the closed-loop system stable. $\mathcal{K}$ is the set which satisfy the \hinf constraint. And the $K^*$ is the controller which minimizes the \htwo cost function. }
\label{fig:cost-function}
  \end{figure}

Let's denote the closed-loop transfer function from inputs to $z_1$ as $T_{\vec{w} \to z_{1}}(s)$ and from the inputs to $z_2$ as $T_{\vec{w} \to z_{2}}(s)$. Then the main goal is to find all the filters, $K=(K_x,K_{\theta},H_x,H_{\theta})$, so that we have minimum horizontal motion on the platform in a Root Mean square (RMS) sense. 

\begin{equation}
K^*=\argmin_{K \in \mathcal{K}} \| T_{\vec{w} \to z_{2}}(s,K) \|_2
\end{equation}

Where $\mathcal{K}$ is the set of all possible filters which satisfy extra constraints which could be enforced for practical reasons. For example, In this work, we need to have a finite upper Unity Gain Frequency (UGF) for both loops. Since the sensor noise exceeds the ground-induced platform motion above a certain frequency, the acausal optimum necessarily coincides with the ground motion. Consequently, to ensure that the closed-loop residual motion on \Ry follows the acausal optimum, the controller must remain disengaged in this frequency range; otherwise, the sensing noise will be injected into the loop, increasing the residual motion beyond the acausal optimum. To enforce a residual motion that closely follows the acausal optimum, we impose the following \hinf constraint on the \Ry loop:

\begin{equation}
\mathcal{K} := \left\{ K \mid \| T_{\vec{w} \to z_{1}}(s,K) \|_{\infty} < \gamma_1 \right\}
\end{equation}

This set can be further restricted by imposing additional requirements such as robustness, phase margins, or stability margins. However, in this work, we focus on a more relaxed constraint set to expose the best achievable performance, serving as a lower bound. Introducing additional constraints beyond those described would likely degrade performance. A graphical representation of the cost function is shown in Figure~\ref{fig:cost-function}.

The $\gamma_1$ is the maximum deviation from the $\xi_{\theta}$ --- which measures how much we are willing to deviate from the acausal optimum. The smaller the $\gamma_1$ the better the performance on \Ry but this would restrict the performance on the x loop. Since we don't have much constraint on the \Ry loop except the finite UGF we can set this value in a relaxed way. Furthermore, instead of enforcing an \hinf\ norm on the \Ry\ loop, it is possible to constrain the unity gain frequency (UGF) by introducing a right-half-plane zero paired with a left-half-plane pole at a similar frequency. These add extra phase loss at high frequencies, enforcing lower unity gain.  This approach shapes the open-loop transfer function to enforce a desired roll-off behavior.

\subsection{Optimization algorithm}

To solve the optimization problem outlined above, several approaches exist. Riccati-based methods are one class of techniques that either directly yield a global optimum or reach it through iteration \cite{doyleStatespaceSolutionsStandard1989, safonovSimplifyingTheoryLoopshifting1989}. However, based on our experience, these methods—and their variants—become numerically unstable for large-scale systems with cross-couplings and nested loops. They also lack flexibility when defining complex cost functions that mix \htwo and \hinf norms. Additionally, these methods typically produce controllers with the same order as the plant, which is not always desirable; lower-order controllers can offer greater robustness and ease of implementation.

In this study, we adopt a subgradient descent method to find optimal solutions \cite{apkarianNonsmoothSynthesis2006, apkarianMultimodelMultiobjectiveTuning2014}. This algorithm is implemented in MATLAB as the \texttt{systune} function. As a gradient-based method, it allows for initialization with informed guesses. We initialize the blending filter using a polynomial blending approach, which accelerates convergence and allows control over filter complexity prior to optimization. Similarly, controller order can be fixed in advance, with optimization applied only to the relevant parameters.

Due to the high dynamic range and small magnitudes typical in gravitational-wave transfer functions, we use acceleration as the transfer function basis and apply nanometer scaling. This reduces the dynamic range, leading to better-conditioned state-space matrices and improved numerical stability.

While this algorithm does not guarantee global optimality, we found it robust in practice: repeated runs with random initializations consistently converged to similar solutions. For each configuration, we performed 30 parallel optimizations using MATLAB’s parallel computing tools. Each batch completed in approximately 5 minutes on a 2020 MacBook Pro (M1, macOS).

\section{Results}

\begin{table}[t]
  \centering
\begin{tabular}{ |p{3cm}||p{1cm}|p{1cm}|p{1cm}|p{1cm}|p{1cm}|  }
  \hline
  \multicolumn{6}{|c|}{Filter size used for optimization } \\
  \hline
  Active platform & $P$ & $K_x$ & $K_{\theta}$ & $H_{x}$ & $H_{\theta}$  \\
  \hline
  OmniSens   & 94  & 6  & 4 &  6 & 1 \\
  T360+BRS   & 84  & 8  & 6 &  7 & 3 \\
  \hline
 \end{tabular}
\caption{Optimization results for both OmniSens and T360+BRS configurations. The reported values indicate the number of states in each system (i.e., the size of the state-space $A$ matrix). $P$ denotes the generalized plant for each sensor setup. $K_x$ and $K_{\theta}$ represent the number of states in the feedback controllers for the $x$ and \Ry\ loops, respectively, while $H_x$ and $H_{\theta}$ correspond to the number of states in the blending filters (including both high-pass and low-pass components).}
\label{tab:optimisation_result}
\end{table}

\begin{figure}[h]

\begin{subfigure}{.48\textwidth}
  \centering
  \includegraphics[width=1\linewidth]{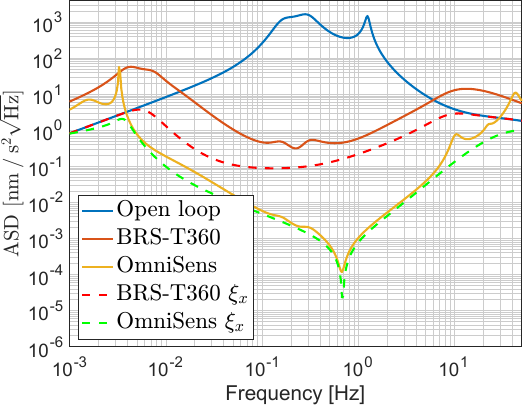}
  \caption{  Acceleration  } 
\end{subfigure}%
\hspace{0.02\textwidth} 
\begin{subfigure}{.48\textwidth}
  \centering
  \includegraphics[width=1\linewidth]{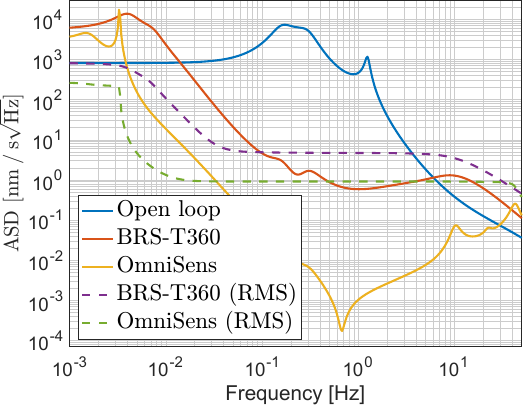}
  \caption{ Velocity }
\end{subfigure}
\caption{Open- and closed-loop horizontal performance comparison between OmniSens and \altsens. The open-loop response is obtained by multiplying the ground model with the ISI ground-to-table transfer function.}
\label{fig:result_comparison}
\end{figure}

We applied the optimization framework described above to both \altsens and OmniSens configurations. Results are presented in Figure~\ref{fig:result_comparison}. For each configuration, the sizes of the controller and blending filters were initialized a priori. Starting from minimal values, we incrementally increased the sizes until no further significant improvement in performance was observed. The results demonstrate that the OmniSens configuration significantly outperforms the BRS-T360, especially near the microseism frequency, reducing noise levels by at least two orders of magnitude. This result is as expected since the OmniSens is sensitive to inertial rotations down to zero frequency and exhibits lower suspension thermal noise. In practice we expect the rotational performance to be limited by direct torque noise acting on the test mass. The final controller and blending filter sizes are summarized in Table~\ref{tab:optimisation_result}. Bode plots of the open-loop transfer functions and blending filters are provided in Appendix~D.

\begin{figure}[ht]
  \centering
\includegraphics[width=0.6\linewidth]{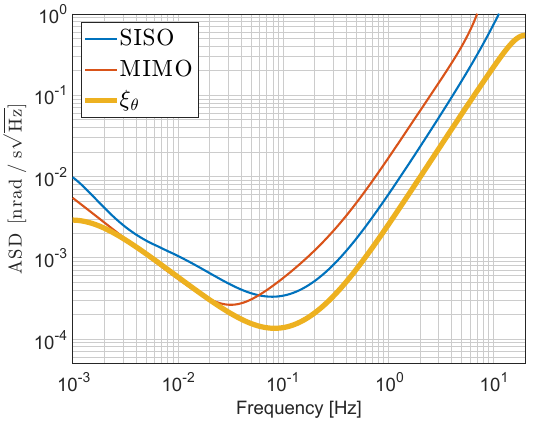}
\caption{ \altsens \Ry loop comparison between two control strategies: the \Ry SISO curve results from optimizing the \Ry loop independently, while the MIMO curve comes from the framework presented here, where \Ry serves as an auxiliary loop and its cross-coupling to the $x$ loop is the primary focus. The MIMO curve places greater emphasis on lower frequencies—where coupling to the $x$ loop is strongest—while deviating from the acausal optimum at higher frequencies.}
\label{fig:altsens-ry-siso-mimo}
\end{figure}

We also evaluated a SISO strategy that optimizes the \Ry loop independently of the x loop. This approach tends to emphasize frequency bands that don't necessarily reduce noise in the degree of freedom of interest. As shown in Figure \ref{fig:altsens-ry-siso-mimo}, the MIMO optimization performs slightly worse at higher frequencies but offers improved performance at lower frequencies. This is primarily due to the significant role of tilt-to-translation coupling in the x loop at low frequencies. All scripts and data supporting this paper are available at Zenodo: \url{https://doi.org/10.5281/zenodo.15830252}. 

\section{Conclusion}

Using the framework presented here, we formulated and solved multi-scale optimal control problems in a consistent and flexible manner. The same methodology extends naturally to more complex systems with additional degrees of freedom—such as the integrated control design for the ET suspension system. The framework's flexibility allows rapid exploration and iteration of different sensor configurations and control strategies. Because suspension parameters can be expressed as free variables in a state-space model, control and mechanical design can be co-optimized within a unified formulation. This makes it possible to fine-tune critical aspects like sensor placement and configuration during early-stage design.    

\section{Acknowledgements}
This project has received funding from the European Research Council (ERC) under the European Union’s Horizon 2020 research and innovation programme (grant agreement No. 865816).

Author Nathan A. Holland has relocated to; LIGO Laboratory, California Institute of Technology, Pasadena, CA 91125, USA.

Author Alexandra L. Mitchell has relocated to; Stanford University, Stanford, CA 94305, USA.

 \clearpage

\appendix

\section{Reference mass model derivation} 

\newcommand{\re}{\operatorname {Re}}
\newcommand{\im}{\operatorname {Im}}
\newcommand{\Lag}{\mathcal{L}}
\newcommand{\half}{\frac{1}{2}}
\newcommand{\xg}{x_{\rm{t}}}
\newcommand{\yg}{y_{\rm{t}}}
\newcommand{\phig}{\phi_{\rm{t}}}
\newcommand{\xm}{x_{\rm{m}}}
\newcommand{\ym}{y_{\rm{m}}}
\newcommand{\phim}{\phi_{\rm{m}}}
\newcommand{\kel}{\kappa_{\rm{el}}}
\newcommand{\dotxm}{\Dot{x}_{\rm{m}}}
\newcommand{\dotym}{\Dot{y}_{\rm{m}}}
\newcommand{\dotphim}{\Dot{\phi}_{\rm{m}}}
\newcommand{\ddotxm}{\Ddot{x}_{\rm{m}}}
\newcommand{\ddotym}{\Ddot{y}_{\rm{m}}}
\newcommand{\ddotphim}{\Ddot{\phi}_{\rm{m}}}
\newcommand{\IM}{I_{\rm{m}}}

Here we will go through the derivation of the equation of motion and subsequently the transfer functions for the OmniSens reference mass which will be used during the studies in this paper.

\subsection{Definitions}

\begin{figure}[htbp]
\begin{center}
\includegraphics[scale=0.7]{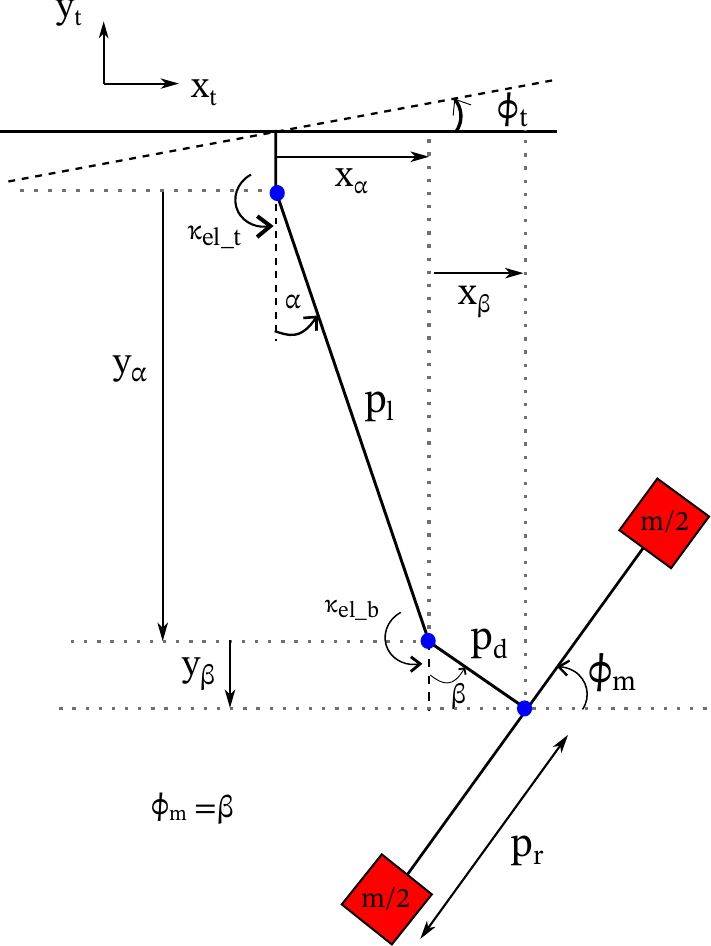}
\caption{Definition of angles, lengths, ground motion and pendulum movements.}
\label{pendulum-diagram}
\end{center}
\end{figure}

Main system diagram which shows the pendulum and coordinate definitions can be seen in figure \ref{pendulum-diagram}.

\begin{table}
\begin{center}
\begin{tabular}{cccc}
 Parameter & Description & Value & Unit \\ \hline
 $g$ & earth's gravitational acceleration & 9.834 & $m/s^{2}$ \\
 $p_{l}$& pendulum's length &  0.7 & $m$  \\
 $p_{d}$ & pendulum's second part length & $-1.8 \times 10^{-4}$ & $m$   \\
 $k_{el_b}$ & bottom spring coefficient & 0.019 & $Kg.m^{2}/s^{2}$ \\
 $m$ & mass & 3.8 & $Kg$ \\
 $I_{m}$ & moment of inertia & 0.95 & $Kg.m^{2}$ \\
 $\phi_{eff}$ & effective loss angle & 1e-6 & rad
\end{tabular}
\caption{Variables and parameters used in the model.}
\label{tab:params}
\end{center}
\end{table}

\subsubsection{Position vectors}

From the diagram we can derive these relationships:

\begin{equation}
x_{m} = x_{\alpha} + x_{\beta} + x_{t}
\end{equation}

\begin{equation}
y_{m} = y_{\alpha} + y_{\beta} + y_{t}
\end{equation}

\begin{equation}
x_{\alpha} = p_l sin(\alpha)
\end{equation}

\begin{equation}
y_{\alpha} = - p_l cos(\alpha)
\end{equation}

\begin{equation}
y_{\beta}=-p_d\cos\beta=-p_d\cos\phim
\end{equation}

At the end we need to some functions which describes the pendulum motion (output)  $\xm$, $\ym$ and $\phim$ in terms of the ground motion (input) $\xg$, $\yg$ and $\phig$. So we need to remove all other time dependent variables by defining them based on our input and output variables.

\begin{equation}
x_{\alpha}=\xm-\xg-x_{\beta}=\xm-\xg-p_d\sin\phim
\end{equation}

\begin{equation}
\alpha=\arcsin\left(\frac{\xm-\xg-p_d\sin\phim}{p_l}\right)
\end{equation}

\subsubsection{Energy variables}

To formulate the lagrangian for the system all of the kinetic and potential energy changes need to be known. 
In this system the energy changes comes from the kinetic energy due to the pendulum motion and the rotation of the mass and the gravitational potential energy and the elastic potential energy due to the bending of the fused silica fibre.

The equation for the kinetic energy due to the pendulum motion is in Equation \ref{eq:Tp}

\begin{equation}\label{eq:Tp}
T_{\rm{p}}=\half m \left(\dotxm^2+\dotym^2\right)
\end{equation}

The equation for the kinetic energy due to the rotation of the mass is given in Equation \ref{eq:Trot}

\begin{equation}\label{eq:Trot}
T_{\rm{rot}}=\half \IM\dotphim^2
\end{equation}

The gravitational potential energy can be expressed as:
x
\begin{equation}\label{eq:Vg}
V_{\rm{g}}=-mg(p_l\cos\alpha+p_d\cos\phim)
\end{equation}

The equation for the elastic potential for bottom bending points is given as follows:

\begin{equation}
V_{\rm{el, bot}}=\half\kel(\phim-\alpha)^2
\end{equation}

The lagrangian will then become:

\begin{equation}\label{eq:Lag1}
\Lag  =T_{\rm{p}}+T_{\rm{rot}}-V_{\rm{g}}-V_{\rm{el_b}} 
\end{equation}

\subsection{Equation of Motion}

The equation of motion (EoM) can be derived from the lagrangian using the Euler-Lagrange equation. We further linearize the EoM around pendulum's rest position. In addition we remove higher order terms which result in a simpler EoM that captures most useful features of the dynamics.

\begin{equation}
\frac{d^{2}}{d t^{2}} x_{m} + (  \frac{g}{p_l}  ) x_{m}  = ( \frac{g}{p_l} ) x_{t}
\end{equation}

\begin{equation}
\frac{d^{2}}{d t^{2}} y_{m} = 0
\end{equation}

\begin{equation}
\frac{d^{2}}{d t^{2}} \phi_{m} + \frac{1}{I_{m}} ( \kappa_{el_b}  + p_dgm  ) \phi_{m} - \frac{1}{I_{m}l} ( \kappa_{el_b} + p_dgm ) x_{m} = -\frac{1}{I_{m}l}  ( \kappa_{el_b} + p_dgm ) x_{t} 
\end{equation}

By using the Laplace transformation we can find the transfer function for the system. The transfer function is defined as the ratio of the output to the input in the Laplace domain.

\begin{equation}
\frac{x_{m}}{x_{t}}(s) = \frac{g}{g + p_l s^{2}}
\end{equation}
 
 \begin{equation}
\frac{\phi_{m}}{x_{t}}(s) = - \frac{s^{2} \left(\kappa_{el_b} + p_d g m\right)}{\left(g + p_l s^{2}\right) \left(I_{m} s^{2} + \kappa_{el_b} + p_d g m\right)}
\end{equation}

Since the sensors are placed on a cage which is connected to the platform, the sensor output is the differential motion of the pendulum with respect to the cage. To find out the transfer function from the platform to the sensor by simple geometry we can write:

 \begin{equation}
\frac{dx}{x_{t}}(s)  = 1 - \frac{x_{m}}{ x_{t}}(s)
\end{equation}
\begin{equation}
\frac{dy2 - dy1}{x_{t}}(s) =  2p_{r} \frac{\phi_{m}}{x_{t}}(s)
\end{equation}
\begin{equation}
\frac{dx}{\phi_{t}}(s) = c_{ls1}
\end{equation}
\begin{equation}
\frac{dy2-dy1}{\phi_{t}}(s) = -2 p_{r}
\end{equation}

\begin{figure}[!t]
  \begin{center}
  \includegraphics[scale=0.7]{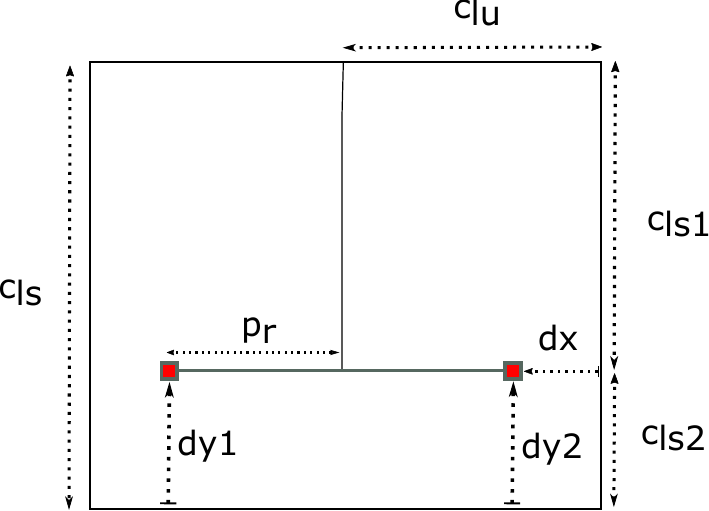}
  \caption{Sensors position relative to the pendulum. }
  \label{fig:sensor-pendulum-diagram}
  \end{center}
  \end{figure}

Where here $dx$ is the horizontal differential sensor output, $dy1$ and $dy2$ are the vertical differential sensors output, $p_{r}$ is pendulum arm length and $c_{ls1}$ the vertical distance of horizontal sensors from the top bending point of the pendulum. For a visual representation of the sensors position relative to the pendulum see figure \ref{fig:sensor-pendulum-diagram}.

\begin{figure*}[h]
  \centering
  \begin{subfigure}[b]{0.48\textwidth}
      \centering
      \includegraphics[width=\textwidth]{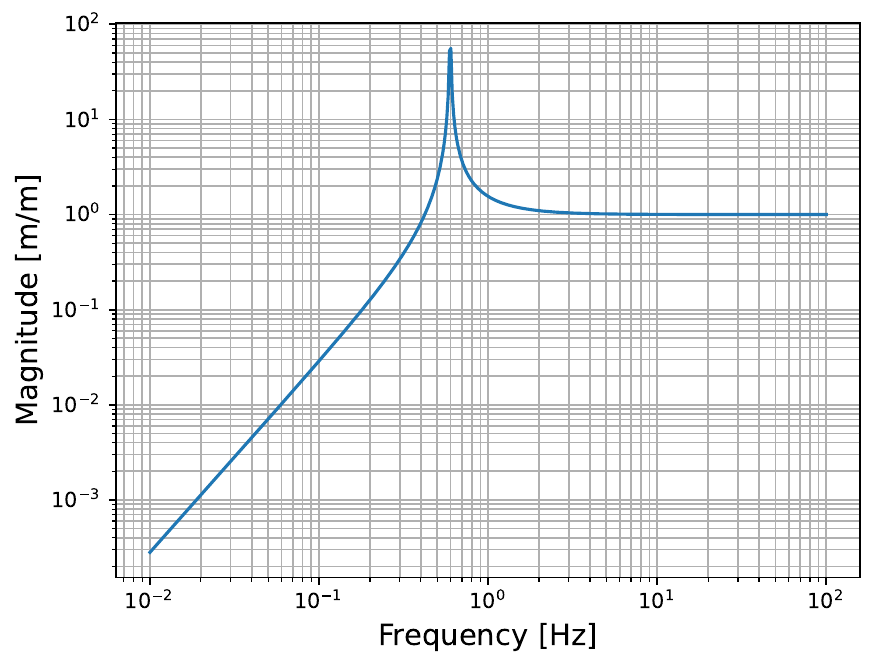}
      \caption[]{Platform translation to sensor horizontal translation}    
      \label{fig:x-x-tf}
  \end{subfigure} 
  \hfill
  \begin{subfigure}[b]{0.48\textwidth}  
      \centering 
      \includegraphics[width=\textwidth]{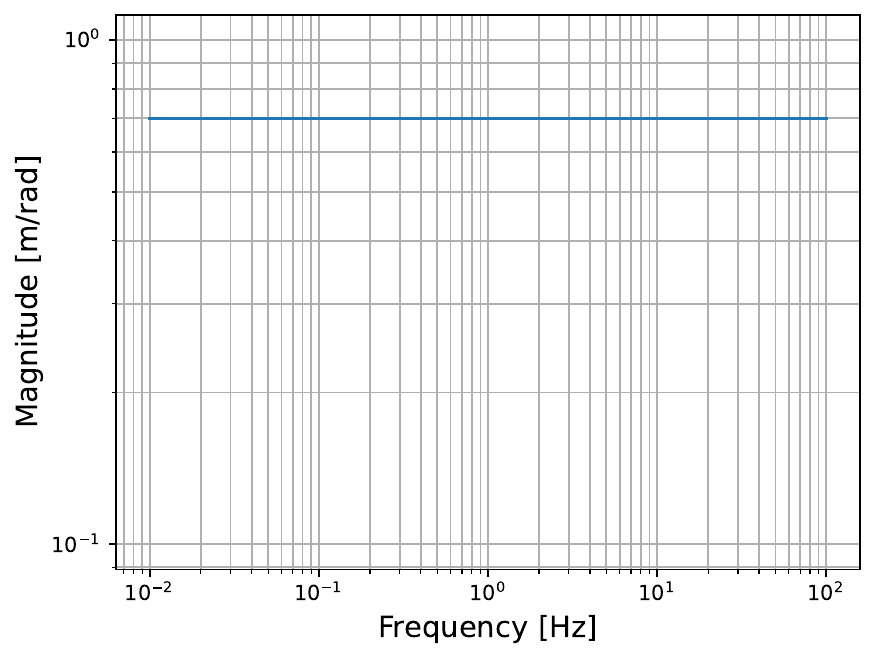}
      \caption[]%
      {Platform rotation to sensor horizontal translation}    
      \label{fig:phi-x-tf}
  \end{subfigure}
  \vskip\baselineskip
  \begin{subfigure}[b]{0.48\textwidth}   
      \centering 
      \includegraphics[width=\textwidth]{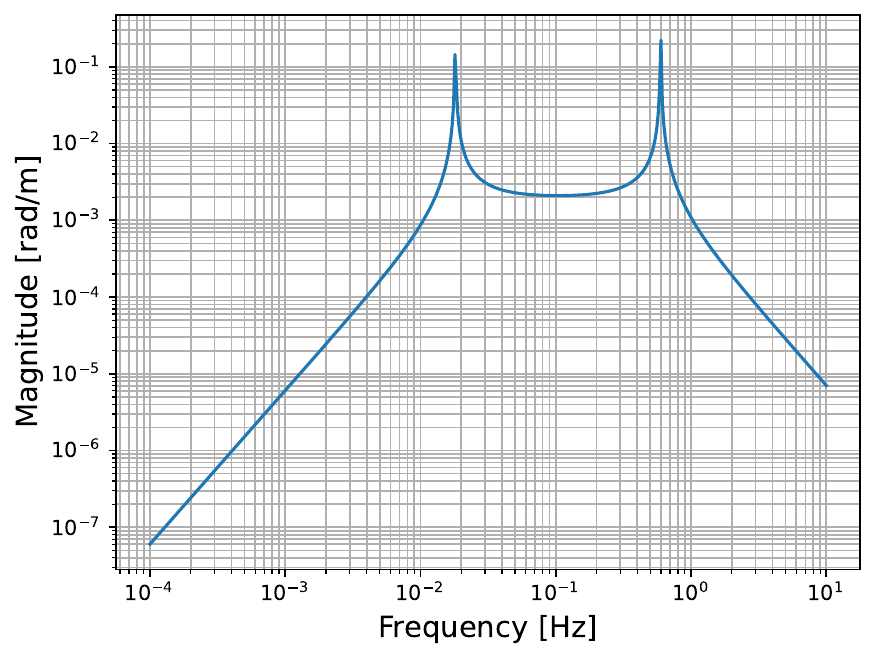}
      \caption[]%
      {Platform translation to sensor rotation}    
      \label{fig:x-phi-tf}
  \end{subfigure}
  \hfill
  \begin{subfigure}[b]{0.48\textwidth}   
      \centering  
      \includegraphics[width=\textwidth]{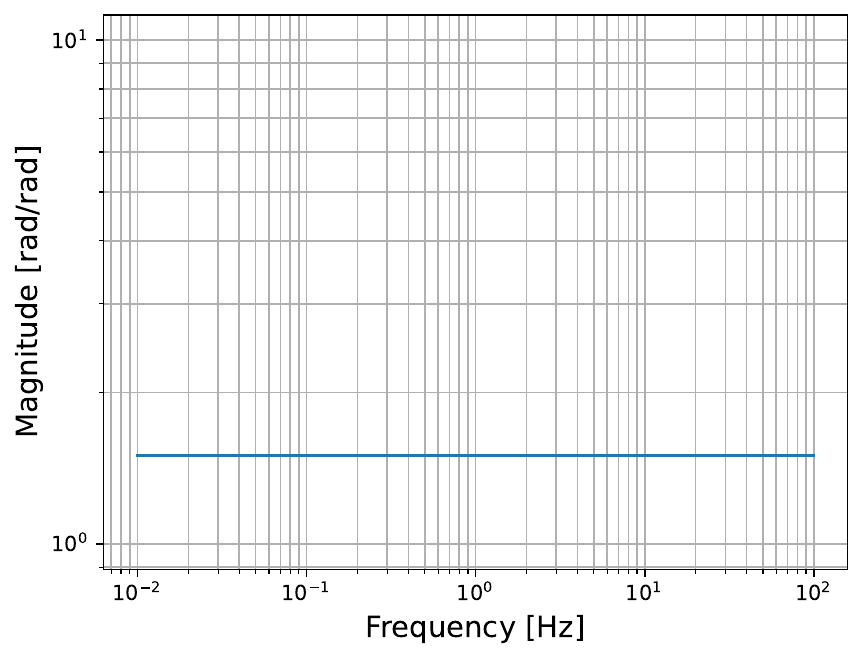}
      \caption[]%
      {Platform rotation to sensor rotation \\}    
      \label{fig:phi-phi-tf}
  \end{subfigure}
  \caption[]
  {(a,b) dx (horizontal sensor) response to input translation $x_{t}$ and rotation $\phi_{t}$ . (c,d) $ dy2-dy1 $ (vertical sensors) response  to input translation $x_{t}$ and rotation $\phi_{t}$.} 
  \label{fig:omnisens-filters}
\end{figure*}

\subsection{Thermal noise in the reference mass}

The thermal noise would contribute to the rotational performance of the pendulum. It acts as a torque on the pendulum which could be modelled as a sensing noise. The power spectral density of the thermal noise is given by \cite{mow-lowry6DInterferometricInertial2019}:

\begin{equation}
S_{torq}(\omega) = 4 K_B T \phi_{eff} k_{el} / \omega
\end{equation}

Where $K_B$ is the Boltzmann constant, $T$ is the room temperature, $\phi_{eff}$ is the effective loss angle of the pendulum and $k_{el}$ is the spring constant. This noise will excite the center of mass of pendulum and result in rotational motion which will be seen by the sensors. To model this we quadrature sum the thermal noise with the sensor noise. This will then be inverted back to $x_t$ and $\phi_{t}$ to get the inertial equivalent noise of the reference mass which will be used in the modeling of the OmniSens configuration.

\section{ISI parameters}

The Internal Seismic Isolation (ISI) for OmniSens is an adapted design of the LIGO Horizontal Access Module ISI (HAM-ISI), as found in \cite{ligo-G070156}. The ISI is a single-stage actively and passively isolated platform. Passive isolation is provided through the suspension, which employs three blade springs (design adapted from \cite{ligo-G070156}) and flexures wires/pendula (design adapted from \cite{ligo-G070156}), manufactured from Titanium Grade-19. Active isolation is achieved through six actuators. The isolated platform is a bolted aluminum rigid hexagonal structure with an optical bench, has a moving mass of $415.35\:\text{kg}$, and a total mass of $625\:\text{kg}$ including payload.

The dynamics of the system are derived from \cite{ligo-G070156}, which return a simplified 6-DoF equation of motion. The model solves the eigenvalue problem in Eq. \eqref{appendix_b_eq_EOM}. The parameters for the OmniSens ISI used in Eq. \eqref{appendix_b_eq_EOM}, are found in Table \ref{tab:appendix_b_tab_parameters}.

\begin{equation}\label{appendix_b_eq_EOM}
    M\mathbf{\ddot{u}} + K \mathbf{u} = \mathbf{0}, \quad \text{where}
\end{equation}

\begin{align}
M &=
    \begin{bmatrix} \notag
        m_{u}& 0 & 0 & 0 & 0 & 0\\
        0 & m_{u} & 0 & 0 & 0 & 0\\
        0 & 0 & m_{u} & 0 & 0 & 0\\
        0 & 0 & 0 & J_{xx} & 0 & 0\\
        0 & 0 & 0 & 0 & J_{yy} & 0\\
        0 & 0 & 0 & 0 & 0 & J_{zz}\\
    \end{bmatrix}, \quad
    \mathbf{u} = 
    \begin{bmatrix} \notag
       x \\
       y \\
       z \\
       \phi \\
       \theta \\
       \psi \\
    \end{bmatrix}, \quad \text{and}\\  
    K &= 
    \begin{bmatrix} \notag
        3k_{xx}& 0 & 0 & 0 & 3k_{xx}h & 0\\
        0 & 3k_{yy} & 0 & 3k_{yy}h & 0 & 0\\
        0 & 0 & 3k_{zz} & 0 & 0 & 0\\
        0 & 3k_{yy}h & 0 & k_{\phi} & 0 & 0\\
        3k_{xx}h & 0 & 0 & 0 & k_{\theta} & 0\\
        0 & 0 & 0 & 0 & 0 & k_{\psi}\\
    \end{bmatrix}
\end{align}

Where here $M$ is the mass matrix, $K$ stiffness matrix, and $u$ is the center of mass position. All parameters are defined in the table \ref{tab:appendix_b_tab_parameters}. The rotational stiffness terms are solved in Eqs. \eqref{appendix_b_eq_platformrotX} - \eqref{appendix_b_eq_platformrotZ}.

\begin{align}
    k_{\phi} =& \frac{3}{2}k_{zz}r_{s}^{2} + 3k_{yy}h^{2} - m_{u}gh - m_{s}gh_{s} \quad \text{rotation around $x$-axis} \label{appendix_b_eq_platformrotX}\\
    k_{\theta} =& \frac{3}{2}k_{zz}r_{s}^{2} + 3k_{xx}h^{2} - m_{u}gh - m_{s}gh_{s} \quad \text{rotation around $y$-axis} \label{appendix_b_eq_platformrotY}\\
    k_{\psi} =& \frac{3}{\sqrt{2}}\sqrt{k_{xx}^{2} + k_{yy}^{2}} r_{s}^2 + 3k_{rz} \quad \text{rotation around $z$-axis} \label{appendix_b_eq_platformrotZ}
\end{align}

\begin{sidewaystable}
\begin{tabular}{llll}
 & \textbf{Parameter} & \textbf{Value} & \textbf{Description} \\ \hline \hline
Platform & $m_{\text{u}}$ & $415.35\:\text{kg}$ & Moving mass of platform \\
 & $m_{\text{s}}$ & $209.65\:\text{kg}$ & Moving mass of payload \\
 & $m$ & $625\:\text{kg}$ & Total mass of platform \\
 & $z_{\text{CoG}}$ & $694.9\:\text{mm}$ & $z$-coordinate of platform center of gravity w.r.t. ground \\
 & $J_{xx}$ & $72.4\:\text{kg}\cdot\text{m}^{2}$ & Moment of inertia around $x$-axis \\
 & $J_{yy}$ & $72.4\:\text{kg}\cdot\text{m}^{2}$ & Moment of inertia around $y$-axis \\
 & $J_{zz}$ & $114.6\:\text{kg}\cdot\text{m}^{2}$ & Moment of inertia around $z$-axis \\
 & $r_{s}$ & $300\:\text{mm}$ & Radius of blade spring tips within platform \\
 & $h$ & $z_{\text{CoG}} - z_{\text{LZMP}}$ & Platform CoG offset from LZMP of flexure \\
 & $z_{\text{s}}$ & $900\:\text{mm}$ & Expected payload CoG w.r.t. ground \\
 & $h_{\text{s}}$ & $z_{\text{s}} - z_{\text{CoG}}$ & Payload CoG offset w.r.t. platform CoG \\ \hline
Blade spring & $k_{zz}$ & $3.34\cdot 10 ^{4}\:\text{N/m}$ & Blade spring translational stiffness along $z$-axis, obtained from FEA \\ \hline
Flexure & $z_{\text{LZMP}}$ & $612.8\:\text{mm}$ & $z$-coordinate of LZMP w.r.t. ground \\
 & $k_{xx}$ & $1.37\cdot10^{4}\:\text{N/m}$ & Flexure translational stiffness along $x$-axis, obtained from FEA \\
 & $k_{yy}$ & $1.37\cdot10^{4}\:\text{N/m}$ & Flexure translational stiffness along $y$-axis, obtained from FEA \\
 & $k_{rz}$ & $3.50\:\text{Nm}\cdot\text{rad}^{-1}$ & Flexure torsional stiffness, obtained from FEA
\end{tabular}
\caption{OmniSens ISI parameters used to solve the equations of of motion} 
\label{tab:appendix_b_tab_parameters}
\end{sidewaystable}

\section{Plant transfer function matrix}

The generalized plant maps input vector $\vec{u}$ to output $\vec{y}$ where these vectors are defined as:

\begin{equation}
  \vec{u} = \begin{bmatrix}
    w_{ \rm \theta ,dist} & w_{ \rm \theta ,sens} & w_{ \rm \theta, disp } & w_{ \rm x, dist} & w_{ \rm x, sens } & w_{ x, disp } & u_1 & u_2 & u_3 & u_4
    \end{bmatrix}^\top
\end{equation}

\begin{equation}
  \vec{y} = 
  \begin{bmatrix}
z_1 & z_2 & y_1 & y_2 & y_3 & y_4
\end{bmatrix}^\top    
\end{equation}

Where both $u$ and $y$ elements are shown in \ref{fig:generalized-plant}. The weighting functions $w_{\theta, p}$ and $w_{x, p}$ can be factored out of generalized plant for cleaner matrix representation:

\begin{equation}
  P = W \odot \tilde{P}
\end{equation}

where $W$ and $P$ denote the weighting function and plant transfer function matrices, respectively. The $\odot$ is the element wise multiplication. The matrices $W$ and $\tilde{P}$ are given by:

\begin{equation}
  \footnotesize
  \tilde{P}(s)=
\begin{bmatrix}
  \eta_{\rm \theta dist} & 0 & 0 & 0 & 0 & 0 & 0 & -p_{\rm \theta} & 0 & 0 \\
  0 & 0 & 0 & \eta_{\rm x, dist} & 0 & 0 & 0 & 0 & 0 & -p_{\rm x} \\
  0 & \eta_{\rm \theta,sens} & -\eta_{\rm \theta,disp} & 0 & 0 & 0 & 0 & 0 & 0 & 0 \\
  \eta_{\rm \theta, dist} & 0 & \eta_{\rm \theta,disp} & 0 & 0 & 0 & 1 & -p_{\rm \theta} & 0 & 0 \\
  T_{\theta \to x} \eta_{\rm{ry,dist}} & 0 & 0 & 0 & \eta_{\rm{x,sens}} & -\eta_{\rm{x,disp}} & 0 & \rm{T_{\theta \to x}} (-p_{\rm{\theta}}) & 0 & 0 \\
  0 & 0 & 0 & \eta_{\rm{x,dist}} & 0 & \eta_{\rm{x,disp}} & 0 & 0 & 1 & -p_{\rm{x}}
\end{bmatrix}
\end{equation}

\begin{equation}
  W(s)=
  \begin{bmatrix}
    w_{\theta , p} & 1 & 1 & 1 & 1 & 1 & 1 & w_{\rm \theta ,p} & 1 & 1 \\
    1 & 1 & 1 & w_{ \rm x,p} & 1 & 1 & 1 & 1 & 1 & w_{ x,p} \\
    1 & 1 & 1 & 1 & 1 & 1 & 1 & 1 & 1 & 1 \\
    1 & 1 & 1 & 1 & 1 & 1 & 1 & 1 & 1 & 1 \\
    1 & 1 & 1 & 1 & 1 & 1 & 1 & 1 & 1 & 1 \\
    1 & 1 & 1 & 1 & 1 & 1 & 1 & 1 & 1 & 1
    \end{bmatrix}    
\end{equation}

\section{Optimized filters}

\subsection{OmniSens filters}

\begin{figure*}[h]
    \centering
    \begin{subfigure}[b]{0.475\textwidth}
        \centering
        \includegraphics[width=\textwidth]{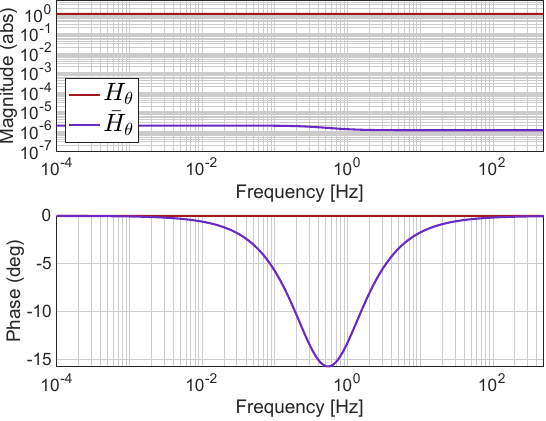}
        \caption[]{Blending filter for \Ry}    
    \end{subfigure}
    \hfill
    \begin{subfigure}[b]{0.475\textwidth}  
        \centering 
        \includegraphics[width=\textwidth]{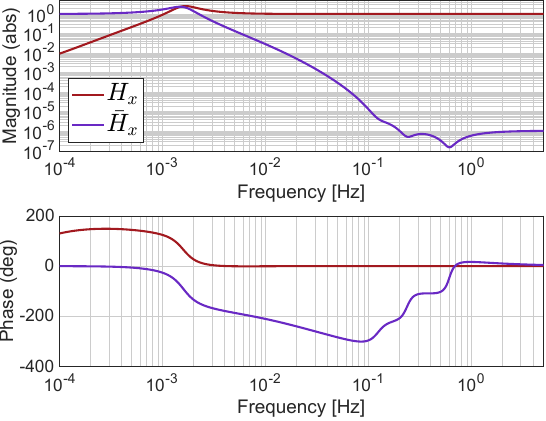}
        \caption[]%
        {Blending filter for X}    
    \end{subfigure}
    \vskip\baselineskip
    \begin{subfigure}[b]{0.475\textwidth}   
        \centering 
        \includegraphics[width=\textwidth]{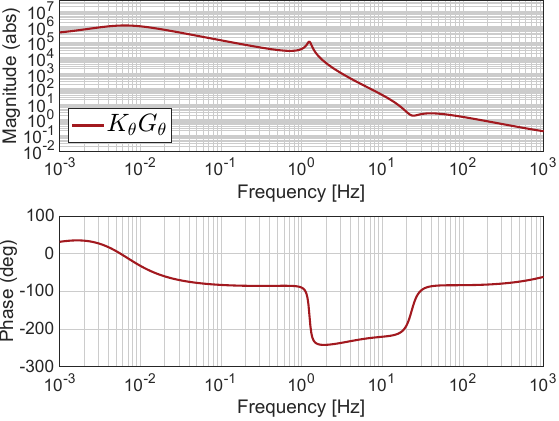}
        \caption[]%
        {Open loop transfer function for \Ry}    
    \end{subfigure}
    \hfill
    \begin{subfigure}[b]{0.475\textwidth}   
        \centering 
        \includegraphics[width=\textwidth]{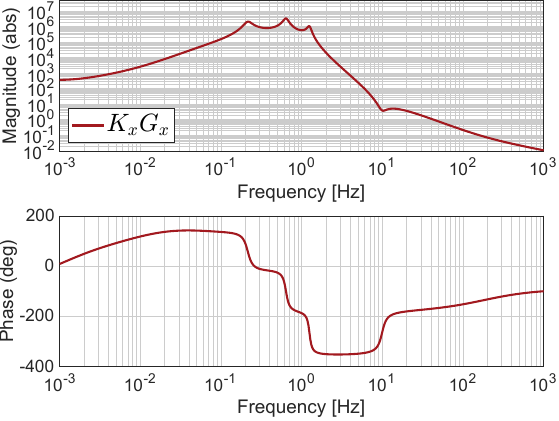}
        \caption[]%
        {open loop transfer function for X}    
    \end{subfigure}
    \caption[]
    {Optimized filters for blending filter and controller for OmniSens. Note that for the \Ry loop the low-pass filter doesn't play any role since the inertial sensor performance exceeds that of the displacement sensor in the table.} 
\end{figure*}

\clearpage

\subsection{\altsens filters}

    \begin{figure*}[h]
      \centering
      \begin{subfigure}[b]{0.475\textwidth}
          \centering
          \includegraphics[width=\textwidth]{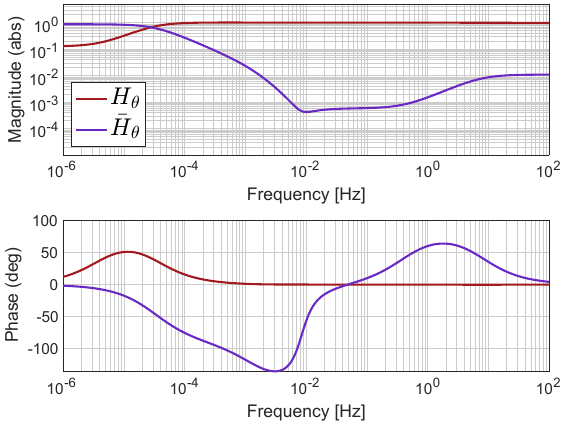}
          \caption[]{Blending filter for \Ry}    
      \end{subfigure}
      \hfill
      \begin{subfigure}[b]{0.475\textwidth}  
          \centering 
          \includegraphics[width=\textwidth]{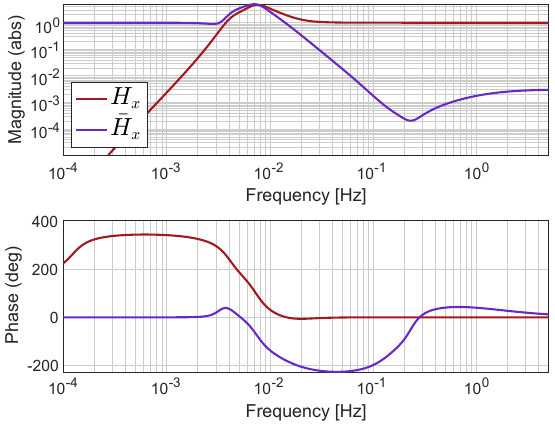}
          \caption[]%
          {Blending filter for x}    
      \end{subfigure}
      \vskip\baselineskip
      \begin{subfigure}[b]{0.475\textwidth}   
          \centering 
          \includegraphics[width=\textwidth]{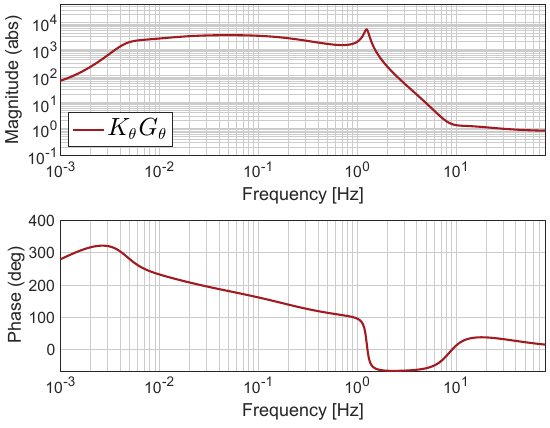}
          \caption[]%
          {Open loop transfer function for \Ry}    
      \end{subfigure}
      \hfill
      \begin{subfigure}[b]{0.475\textwidth}   
          \centering 
          \includegraphics[width=\textwidth]{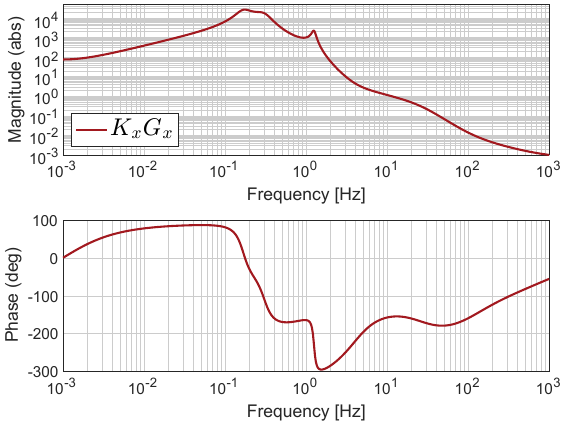}
          \caption[]%
          {open loop transfer function for x}    
      \end{subfigure}
      \caption[]
      {Optimized filters for blending filter and controller for \altsens .} 
      \label{fig:altsens-filters}
  \end{figure*}

\bibliographystyle{unsrt}  
\bibliography{main}  

\begin{thebibliography}{10}

\bibitem{collaborationAdvancedLIGO2015}
The LIGO~Scientific Collaboration et~al.
\newblock Advanced {{LIGO}}.
\newblock {\em Classical and Quantum Gravity}, 32(7):074001, March 2015.

\bibitem{acerneseAdvancedVirgoSecondgeneration2014}
F~Acernese et~al.
\newblock Advanced {{Virgo}}: A second-generation interferometric gravitational
  wave detector.
\newblock {\em Classical and Quantum Gravity}, 32(2):024001, December 2014.

\bibitem{capoteAdvancedLIGODetector2025}
E.~Capote et~al.
\newblock Advanced {{LIGO}} detector performance in the fourth observing run.
\newblock 111(6):062002.

\bibitem{acerneseVirgoDetectorCharacterization2023}
F~Acernese et~al.
\newblock Virgo detector characterization and data quality: Results from the
  {{O3}} run.
\newblock 40(18):185006.

\bibitem{matichardReviewTiltFreeLowNoise2015}
Fabrice Matichard and Matthew Evans.
\newblock Review: {{Tilt}}-{{Free Low}}-{{Noise Seismometry}}.
\newblock {\em Bulletin of the Seismological Society of America},
  105(2A):497--510, April 2015.

\bibitem{maggioreScienceCaseEinstein2020}
Michele Maggiore, Chris Van~Den Broeck, Nicola Bartolo, Enis Belgacem, Daniele
  Bertacca, Marie~Anne Bizouard, Marica Branchesi, Sebastien Clesse, Stefano
  Foffa, Juan {Garc{\'i}a-Bellido}, Stefan Grimm, Jan Harms, Tanja Hinderer,
  Sabino Matarrese, Cristiano Palomba, Marco Peloso, Angelo Ricciardone, and
  Mairi Sakellariadou.
\newblock Science case for the {{Einstein}} telescope.
\newblock {\em Journal of Cosmology and Astroparticle Physics}, 2020(03):050,
  March 2020.

\bibitem{punturoEinsteinTelescopeThirdgeneration2010}
M~Punturo et~al.
\newblock The {{Einstein Telescope}}: A third-generation gravitational wave
  observatory.
\newblock {\em Classical and Quantum Gravity}, 27(19):194002, September 2010.

\bibitem{hildSensitivityStudiesThirdgeneration2011}
S~Hild et~al.
\newblock Sensitivity studies for third-generation gravitational wave
  observatories.
\newblock {\em Classical and Quantum Gravity}, 28(9):094013, April 2011.

\bibitem{mow-lowry6DInterferometricInertial2019}
Conor~M. {Mow-Lowry} and Denis Martynov.
\newblock A {{6D}} interferometric inertial isolation system.
\newblock {\em Classical and Quantum Gravity}, 36(24):245006, December 2019.

\bibitem{Thomas2019}
Dehaeze Thomas, Verma Mohit, and Collette Christophe.
\newblock Complementary filters shaping using $h_{\infty}$ synthesis.
\newblock {\em 2019 7th International Conference on Control, Mechatronics and
  Automation (ICCMA)}, pages 459--464, 2019.

\bibitem{tsangOptimizingActiveSeismic2025}
Terrence Tsang, Fabián Erasmo~Peña Arellano, Takafumi Ushiba, Ryutaro
  Takahashi, Yoichi Aso, and Katherine Dooley.
\newblock Optimizing {{Active Seismic Isolation Systems}} in
  {{Gravitational-Wave Detectors}}.

\bibitem{Sander2025}
Sander~K. Sijtsma, Pooya Saffarieh, Nathan~A. Holland, Sil~T. Spanjer, Wouter
  B.~J. Hakvoort, and Conor~M. Mow-Lowry.
\newblock Non-{{Smooth Multi-objective Controller Synthesis}} for {{Test-Mass
  Actuation}} in {{Gravitational-Wave Detectors}}.

\bibitem{T360}
{Nanometrics}.
\newblock {\em {Trillium} 360 {GSN}}.
\newblock Nanometrics, 3001 Solandt Road, Kanata, Ontario, Canada K2K 2M8,
  1001.18.08 edition, 2023.

\bibitem{venkateswaraHighprecisionMechanicalAbsoluterotation2014}
Krishna Venkateswara, Charles~A. Hagedorn, Matthew~D. Turner, Trevor Arp, and
  Jens~H. Gundlach.
\newblock A high-precision mechanical absolute-rotation sensor.
\newblock {\em Review of Scientific Instruments}, 85(1):015005, January 2014.

\bibitem{mitchellIntegrationHighperformanceCompact2024}
Alexandra Mitchell, Johannes Lehmann, Philip Koch, Samuel Cooper, Jesse van
  Dongen, Leonid Prokhorov, Nathan Holland, Michele Valentini, and Conor
  {Mow-Lowry}.
\newblock Integration of high-performance compact interferometric sensors in a
  suspended interferometer, September 2024.

\bibitem{LIGO_G070156}
{LIGO Scientific Collaboration}.
\newblock {LIGO-G070156: Technical Document}, 2007.
\newblock Accessed: 2025-05-09.

\bibitem{INGV_RSN_2005}
{Istituto Nazionale di Geofisica e Vulcanologia (INGV)}.
\newblock {Rete Sismica Nazionale (RSN)}, 2005.
\newblock Dataset.

\bibitem{LIGO_BRS_TiltSubtraction_2017}
Krishna Venkateswara, Charles~A. Hagedorn, Jens~H. Gundlach, Jeffery Kissel,
  Jim Warner, Hugh Radkins, Thomas Shaffer, Brian Lantz, Richard Mittleman,
  Fabrice Matichard, and Robert Schofield.
\newblock Subtracting tilt from a horizontal seismometer using a
  ground‐rotation sensor.
\newblock {\em Bulletin of the Seismological Society of America},
  107(2):709--717, 03 2017.

\bibitem{Windproofing_LIGO_2020}
Michael~P Ross, Krishna Venkateswara, Conor Mow-Lowry, Sam Cooper, Jim Warner,
  Brian Lantz, Jeffrey Kissel, Hugh Radkins, Thomas Shaffer, Richard Mittleman,
  Arnaud Pele, and Jens Gundlach.
\newblock Towards windproofing ligo: reducing the effect of wind-driven floor
  tilt by using rotation sensors in active seismic isolation.
\newblock {\em Classical and Quantum Gravity}, 37(18):185018, aug 2020.

\bibitem{RossMP_PhDthesis_2020}
Michael~P Ross.
\newblock {\em Precision Mechanical Rotation Sensors for Terrestrial
  Gravitational Wave Observatories}.
\newblock PhD thesis, University of Washington, 2020.

\bibitem{rossVacuumcompatibleCylindricalInertial2023}
M.~P. Ross, J.~{van Dongen}, Y.~Huang, H.~Zhou, Y.~Chowdhury, S.~K. Apple,
  C.~M. {Mow-Lowry}, A.~L. Mitchell, N.~A. Holland, B.~Lantz, E.~Bonilla,
  A.~Engl, A.~Pele, D.~Griffith, E.~Sanchez, E.~A. Shaw, C.~Gettings, and
  J.~Gundlach.
\newblock A vacuum-compatible cylindrical inertial rotation sensor with
  picoradian sensitivity.
\newblock {\em Review of Scientific Instruments}, 94(9):094503, September 2023.

\bibitem{zhouEssentialsRobustControl1998}
Kemin Zhou and John~Comstock Doyle.
\newblock {\em Essentials of Robust Control}, volume 104.
\newblock Prentice hall Upper Saddle River, NJ.

\bibitem{oustaloupFrequencybandComplexNoninteger2000}
A.~Oustaloup, F.~Levron, B.~Mathieu, and F.M. Nanot.
\newblock Frequency-band complex noninteger differentiator: Characterization
  and synthesis.
\newblock 47(1):25--39.

\bibitem{doyleStatespaceSolutionsStandard1989}
J.C. Doyle, K.~Glover, P.P. Khargonekar, and B.A. Francis.
\newblock State-space solutions to standard $h_2$ and $h_{\infty}$ control
  problems.
\newblock {\em IEEE Transactions on Automatic Control}, 34(8):831--847, August
  1989.

\bibitem{safonovSimplifyingTheoryLoopshifting1989}
Michael~G. Safonov et~al.
\newblock Simplifying the $h_{\infty}$ theory via loop-shifting, matrix-pencil
  and descriptor concepts.
\newblock {\em International Journal of Control}, 50(6):2467--2488, December
  1989.

\bibitem{apkarianNonsmoothSynthesis2006}
P.~Apkarian and D.~Noll.
\newblock Nonsmooth $h_{\infty}$ {{Synthesis}}.
\newblock {\em IEEE Transactions on Automatic Control}, 51(1):71--86, January
  2006.

\bibitem{apkarianMultimodelMultiobjectiveTuning2014}
Pierre Apkarian, Pascal Gahinet, and Craig Buhr.
\newblock Multi-model, multi-objective tuning of fixed-structure controllers.
\newblock In {\em 2014 {{European Control Conference}} ({{ECC}})}, pages
  856--861, June 2014.

\bibitem{ligo-G070156}
Brian Lantz.
\newblock Advanced ligo single stage ham (isi) vibration isolation table (final
  design review, fdr).
\newblock \url{https://dcc.ligo.org/LIGO-G070156-v1}.
\newblock LIGO-G070156-v1, LIGO Scientific Collaboration.

\end{thebibliography}

\end{document}